# Performance Analysis of Model-Free Adaptive Control

Feilong Zhang

***Abstract*—We analyzed the model-free adaptive control (MFAC) through the closed-loop function to study its essence and the correct way of reproposing it.**

## I. Introduction

A significant number of works on MFAC have been published in the past decade. We analyzed this method using a closed-loop function and obtained results that differ from those presented in three representative works on current MFAC [1]-[3].

i) The sign of the estimated leading coefficient of the control input is restricted to be unchangeable in the controller [1]-[3]. However, the leading coefficient in the actual model may change, causing the failure of the controller. Formula (13) in [1] lets $\hat{\phi}_{f,L_y,L_u}(k)=\hat{\phi}_{f,L_y,L_u}(1)$ if $sign(\hat{\phi}_{f,L_y,L_u}(k))\neq sign(\hat{\phi}_{f,L_y,L_u}(1))$, and this is the necessary precondition for the system stability analysis through the current contraction mapping technique [1]-[3]. Consequently, this restriction hinders the established model, namely the equivalent-dynamic-linearization model (EDLM), from accurately reflecting the actual system, potentially leading to controller failure. We have discussed this problem in Example 1.

ii) The static error of the speed response of the system is eliminated when $\lambda$=0, as proven in this brief. This conclusion differs from [1]-[3], which showed that the tracking error of the system controlled by MFAC converges to zero when $\lambda$ is large enough. Furthermore, when the desired trajectory is a step signal, the reason behind the convergence of the tracking error to zero in the MFAC-controlled system is the inherent presence of an integrator in the current MFAC rather than being determined by $\lambda$.

iii) The orders of function (1) in [1] should not be $n_y$, $n_u$ but $n_y$+1, $n_u$+1 [6]. The two pseudo orders $L_y$ and $L_u$ in current works are limited to $1\le L_y\le n_y$ and $1\le L_u\le n_u$, respectively. However, according to [4]-[6], the most important and ideal choice of pseudo gradient orders is the actual gradient orders $L_y=n_y+1$ and $L_u=n_u+1$ in adaptive control. To this end, we extend the range to $0\le L_y$ and $1\le L_u$.

Furthermore, we have analyzed the MFAC for the multivariable systems in [7]-[9].

On the other hand, some noteworthy merits of the proposed method are shown as follows.

Manuscript received Dec 3, 2020. This work was supported in part by the xxxxxxxxxxxxx.

Feilong Zhang is with the State Key Laboratory of Robotics, Shenyang Institute of Automation, Chinese Academy of Sciences, Shenyang 110016, China (e-mail: zhangfeiong@sia.cn).

[10]-[14] decompose the nonlinear auto regressive and moving average model (NARMAM) or nonlinear industrial process model into a simple linear model and an unmolded dynamics (UD) around an operating point. Then the corresponding controller with compensation of UD is designed and analyzed using the closed-loop system equation. In comparison to previous works, our proposed equivalent-dynamic-linearization model (EDLM) offers the advantage of describing NARMAM at any point, making it easier and more widely applicable for MFAC. On the other hand, can anyone achieve the tracking performance of Example 3 within 15 minutes? After mastering the proposed method, one can easily achieve that.

The remainder of this paper is organized as follows: Section II presents the corrected EDLM and MFAC. Then the stability of the system controlled by MFAC is analyzed through the closed-loop system function and the simulations are presented. Section III studies how to apply MFAC in nonlinear systems. Section IV gives the conclusion.

## II. Equivalent Dynamic Linearization Model and Design of Model-Free Adaptive Control

In Part A of this section, we correct the current EDLM and reintroduce its fundamental assumptions and theorem. In Part B, we design the MFAC controller and analyze its performance.

### A. *Equivalent Dynamic Linearization Model*

We consider the following discrete-time SISO system:

$$y(k+1)=f(y(k),\cdots,y(k-n_y),u(k),\cdots,u(k-n_u)) \quad (1)$$

where $f(\cdots)\in R$ represents the nonlinear function; $u(k)$ and $y(k)$ denote the input and output of the system at time $k$, respectively. And $n_y$+1, $n_u$+1$\in$Z represent their orders.

Let

$$\boldsymbol{\varphi}(k)=[y(k),\cdots,y(k-n_y),u(k),\cdots,u(k-n_u)]^T \quad (2)$$

Then (1) can be rewritten as

$$y(k+1)=f(\boldsymbol{\varphi}(k)) \quad (3)$$

*Assumption 1*: The partial derivatives of $f(\cdots)$ with respect to all variables are continuous.

*Theorem 1*: Given the nonlinear system (1) satisfying *Assumptions 1*, if $\Delta\boldsymbol{H}(k)\neq 0$, there is a time-varying vector named PG vector, and the system (1) can be transformed into the following full-form equivalent-dynamic -linearization model

$$\Delta y(k+1)=\boldsymbol{\phi}_L^T(k)\Delta\boldsymbol{H}(k) \quad (4)$$

where

$$\boldsymbol{\phi}_L(k)=\begin{bmatrix}\boldsymbol{\phi}_{Ly}(k)\\ \boldsymbol{\phi}_{Lu}(k)\end{bmatrix}=[\phi_1(k),\cdots,\phi_{Ly}(k),\phi_{Ly+1}(k),\cdots,\phi_{Ly+Lu}(k)]^T ,$$

and $\Delta \boldsymbol{H}(k)=\begin{bmatrix}\Delta \boldsymbol{Y}_{Ly}(k)\\ \Delta \boldsymbol{U}_{Lu}(k)\end{bmatrix}=[\Delta y(k),\cdots,\Delta y(k-L_y+1), \Delta u(k),\cdots,\Delta u(k-L_u+1)]^T$ is a vector that consists of an increment of control input within the time window $[k-L_u+1,k]$ and an increment of system output within the time window $[k-L_y+1,k]$. Two integers $0\le L_y$, $1\le L_u$ are named pseudo orders of the system.

And we define $\boldsymbol{\phi}_{Ly}(z^{-1})=\phi_1(k)+\cdots+\phi_{Ly}(k)z^{-Ly+1}$, $\boldsymbol{\phi}_{Lu}(z^{-1})=\phi_{Ly+1}(k)+\cdots+\phi_{Ly+Lu}(k)z^{-Lu+1}$, and $z^{-1}$ is the backward-shift operator.

*Remark 1*: [1] and [2] give the proof of *Theorem 1* in the case of $1\le L_y\le n_y$, $1\le L_u\le n_u$ and we further prove *Theorem 1* for the orders $0\le L_y$ and $1\le L_u$ in Appendix.

We prefer $L_y=n_y+1$ and $L_u=n_u+1$ in applications if $n_y$ and $n_u$ can be obtained. Otherwise, we usually choose the proper $L_y$ and $L_u$ that satisfy $n_y+1\le L_y$ and $n_u+1\le L_u$ in adaptive control. One reason is that the online estimated coefficients of redundant items $\Delta y(k-n_y-1),\cdots,\Delta y(k-L_y+1)$ and $\Delta u(k-n_u-1),\cdots,\Delta u(k-L_u+1)$ might be close to zero, and simultaneously the estimated coefficients of $\Delta y(k),\cdots,\Delta y(k-n_y),\Delta u(k),\cdots,\Delta u(k-n_u)$ will be closer to the actual values compared to $0\le L_y\le n_y$ and $1\le L_u\le n_u$.

### *B. Design of Model Free Adaptive Control*

We can rewrite (4) into (5).

$$y(k+1)=y(k)+\boldsymbol{\phi}_L^T(k)\Delta \boldsymbol{H}(k) \tag{5}$$

The object is to design a controller that optimizes output tracking performance in the sense that:

$$J=\left|y^*(k+1)-y(k+1)\right|^2=\min imum \tag{6}$$

where $y^*(k+1)$ is the desired system output signal.

Substitute Equation (5) into Equation (6) and solve the optimization condition $\partial J/\partial \Delta u(k)=0$, then we have:

$$\begin{aligned}\Delta u(k)=&\frac{1}{\phi_{Ly+1}(k)}[y^*(k+1)-y(k)-\sum_{i=1}^{Ly}\phi_i(k)\Delta y(k-i+1)\\&-\sum_{i=Ly+2}^{Ly+Lu}\phi_i(k)\Delta u(k+L_y-i+1)]\end{aligned} \tag{7}$$

Herein, we change the coefficient $\frac{1}{\phi_{Ly+1}(k)}$ into $\frac{\phi_{Ly+1}(k)}{\lambda(k)+\phi_{Ly+1}^2(k)}$ to prevent the denominator from being zero and let $\lambda(k)=\lambda$ for easier performance analysis. Then the controller will become (8).

$$\begin{aligned}\Delta u(k)=&\frac{\phi_{Ly+1}(k)}{\lambda+\phi_{Ly+1}^2(k)}[y^*(k+1)-y(k)-\sum_{i=1}^{Ly}\phi_i(k)\Delta y(k-i+1)\\&-\sum_{i=Ly+2}^{Ly+Lu}\phi_i(k)\Delta u(k+L_y-i+1)]\end{aligned} \tag{8}$$

According to [1], [2], (8) is also the optimal solution of (9).

$$J=\left|y^*(k+1)-y(k+1)\right|^2+\lambda\left|\Delta u(k)\right|^2 \tag{9}$$

Form (4) and (8), we can have

$$\begin{aligned}&\left[\lambda(1-z^{-1})\left[1-z^{-1}\boldsymbol{\phi}_{Ly}(z^{-1})\right]+\phi_{Ly+1}(k)\boldsymbol{\phi}_{Lu}(z^{-1})\right]y(k+1)\\&=\phi_{Ly+1}(k)\boldsymbol{\phi}_{Lu}(z^{-1})y^*(k+1)\end{aligned} \tag{10}$$

The function of the closed-loop poles is

$$T(z^{-1})=\lambda(1-z^{-1})\left[1-z^{-1}\boldsymbol{\phi}_{Ly}(z^{-1})\right]+\phi_{Ly+1}(k)\boldsymbol{\phi}_{Lu}(z^{-1}) \tag{11}$$

We may place the closed-loop poles in the unit circle to guarantee the system stability by tuning $\lambda$ and quantitatively analyze the chosen $\lambda$ based on the locations of closed-loop poles.

The steady-state error is typically calculated for linear systems. In this paper, we regard the steady state error as one property of the transient tendency of the nonlinear systems. It means the tracking error of nonlinear system (1) described by linear system (5) at the time $k$ trends to. The "steady state error" in the speed (ramp) response of the nonlinear system, as described by the linear system model (5), at time $k$ is

$$\begin{aligned}\lim_{k\to\infty}e(k)&=\lim_{z\to1}\frac{z-1}{z}\left(1-\frac{\phi_{Ly+1}(k)\boldsymbol{\phi}_{Lu}(z^{-1})}{T(z^{-1})}\right)\frac{T_s z}{(z-1)^2}\\&=\lim_{z\to1}\left(\frac{\lambda T_s\left[1-\boldsymbol{\phi}_{Ly}(z^{-1})\right]}{\lambda(1-z^{-1})\left[1-\boldsymbol{\phi}_{Ly}(z^{-1})\right]+\phi_{Ly+1}(k)\boldsymbol{\phi}_{Lu}(z^{-1})}\right)\end{aligned} \tag{12}$$

where $T_s$ represents the sample time constant. We can conclude that the "steady state error" in the speed response positively relates to $\lambda$. Specifically, when $\lambda=0$, we will have $\lim_{k\to\infty}e(k)=0$.

This conclusion differs from the theorem that the convergence of tracking error of the system controlled by MFAC is guaranteed under the condition that $\lambda$ is large enough in [1]-[3]. Besides, the tracking error of step response of the system controlled by the current MFAC converges to zero can be ascribed to that the current MFAC naturally contains one integrator and is unrelated to $\lambda$.

Furthermore, when system stability can be ensured by setting $\lambda=0$, the static error will be eliminated for a desired trajectory of $k^n$ $(0\le n<\infty)$ since

$$\begin{aligned}\lim_{k\to\infty}e(k)&=\lim_{\substack{z\to1\\ \lambda=0}}\frac{z-1}{z}\left(1-\frac{\phi_{Ly+1}(k)\boldsymbol{\phi}_{Lu}(z^{-1})}{T(z^{-1})}\right)\frac{C(z)}{(z-1)^{n+1}}\\&=\lim_{\substack{z\to1\\ \lambda=0}}\left(\frac{\lambda\left[1-\boldsymbol{\phi}_{Ly}(z^{-1})\right]}{\lambda(1-z^{-1})\left[1-\boldsymbol{\phi}_{Ly}(z^{-1})\right]+\phi_{Ly+1}(k)\boldsymbol{\phi}_{Lu}(z^{-1})}\right)\frac{C(z)}{(z-1)^{n-1}}\\&=0\end{aligned} \tag{13}$$

where $Z(k^n)=\frac{C(z)}{(z-1)^{n+1}}$, . $C(z)$ .is the polynomial with the highest power of $n$ and $Z(\bullet)$ denotes $z$-transformation.

*Simulations:*

*Example* 1: The following discrete-time SISO structure-varying linear system is considered in this example.

$$y(k+1)=\begin{cases}-0.4y(k)-0.5u(k)-0.6u(k-1)+d_1 & 1\le k\le 350\\ 0.4y(k)+0.5u(k)+0.6u(k-1)+d_2 & 351\le k\le 700\end{cases} \tag{14}$$

where $d$ is the disturbance. The desired output trajectory is

$$y^*(k+1)=\begin{cases}0.4^{round(k/50)} & 1\le k\le 490\\ 0.1+0.1\times(-1)^{round(k/50)} & 491\le k\le 700\end{cases}$$

The controller parameters and initial values for MFAC are listed in Table I. The estimation algorithm adopts the projection algorithm in [1], and [2] with tuning parameters $\eta$ and $\mu$.

TABLE I Parameter Settings for MFAC

| Parameter | MFAC (6) |
|---|---|
| Order | $L_y=1,\ L_u=2$ |
| $\eta$; $\mu$; $\lambda$ | 3; 1; 0.2 |
| Initial value $\hat{\phi}_L(1)$ | [-0.1, -0.1, -0.1] |
| $u(0:6)$ | $(0,0,0,0,0,0)$ |
| $y(0:5)$ | $(0,0,0,0.5,0.2)$ |

Case 1, $d_1$ =1 and $d_2$ =100.

Fig. 1 shows the tracking performance of the system controlled by MFAC. Fig. 2 shows the control input. Fig. 3 shows the elements in the estimated PG vector.

From Fig. 1, we can see that the MFAC can remove the influence of constant disturbance on the static error of the system. Because this kind of controller inherently contains one integrator.

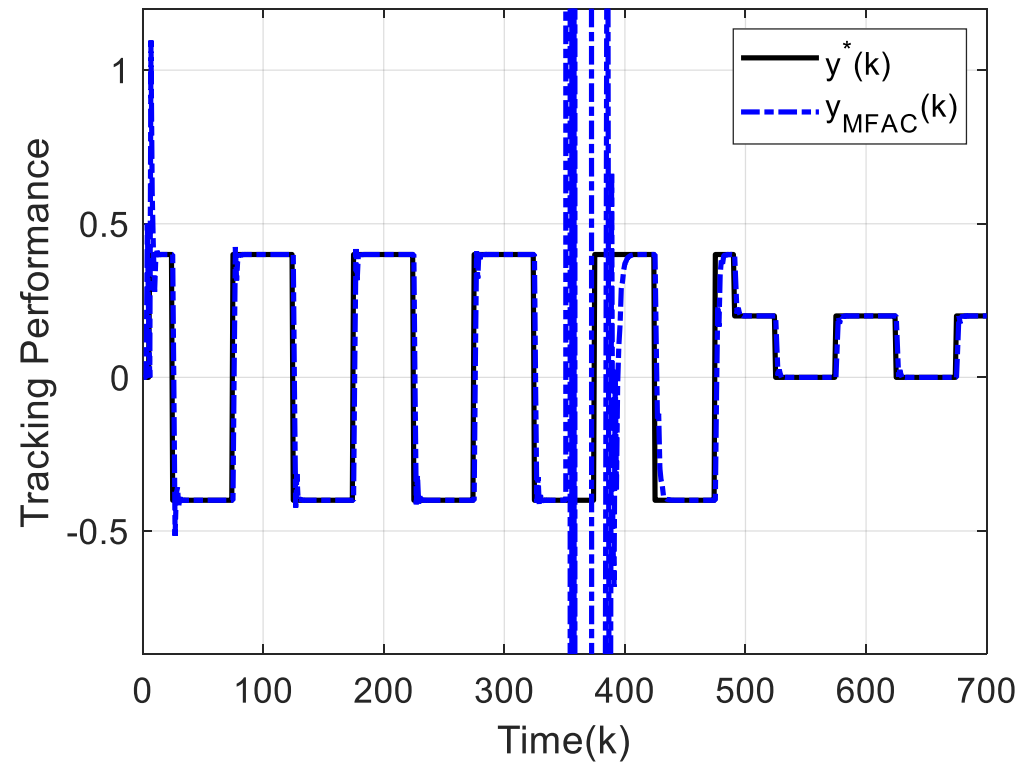

Fig. 1 Tracking performance

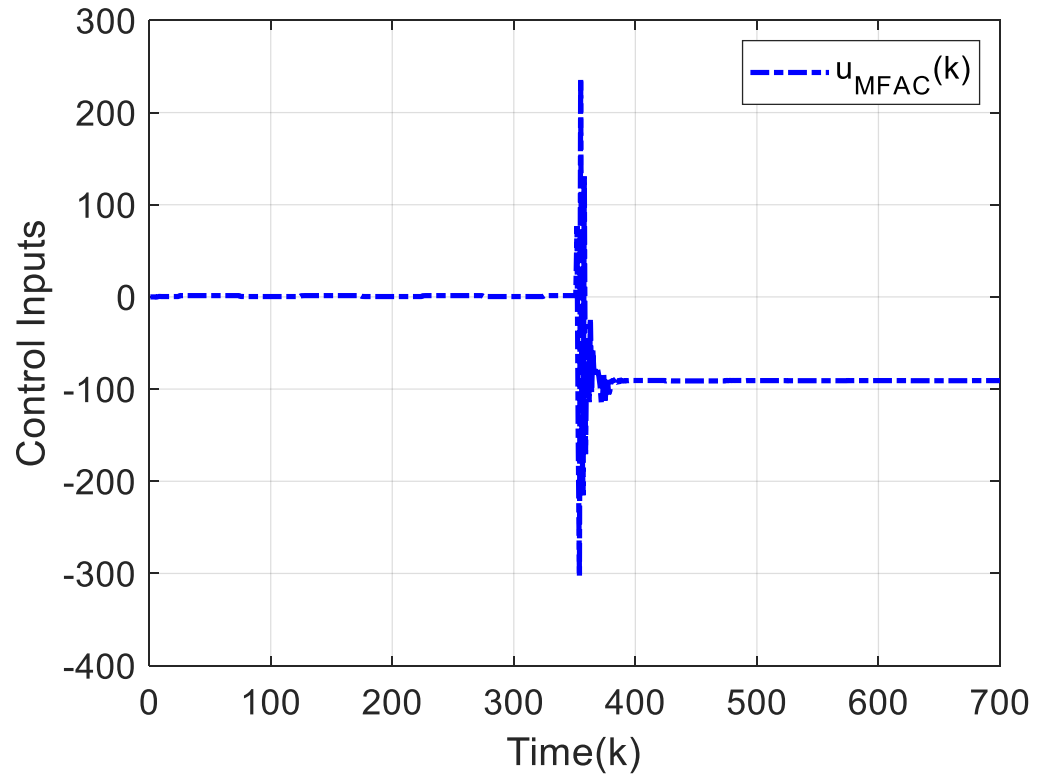

Fig. 2 Control input

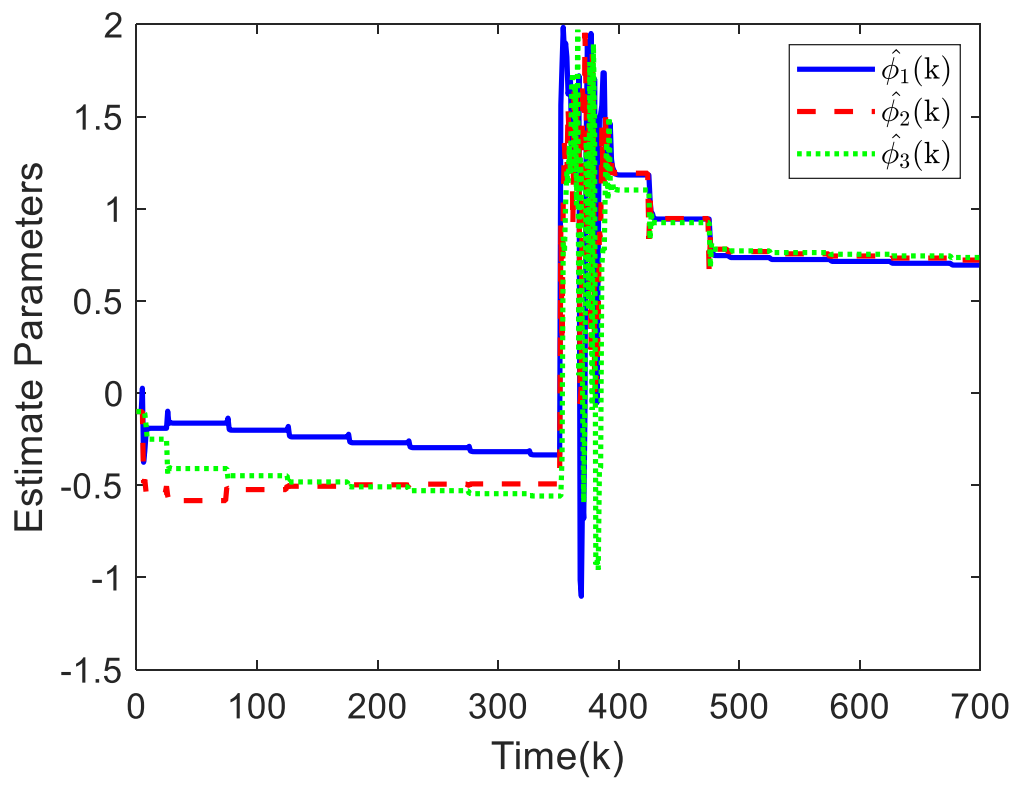

Fig. 3 Elements in estimated PG vector

Case 2, $d_1$=0 and $d_2$=0.

Fig. 4 shows the tracking performance of the system controlled by MFAC. Fig. 5 shows the elements in the estimated PG vector.

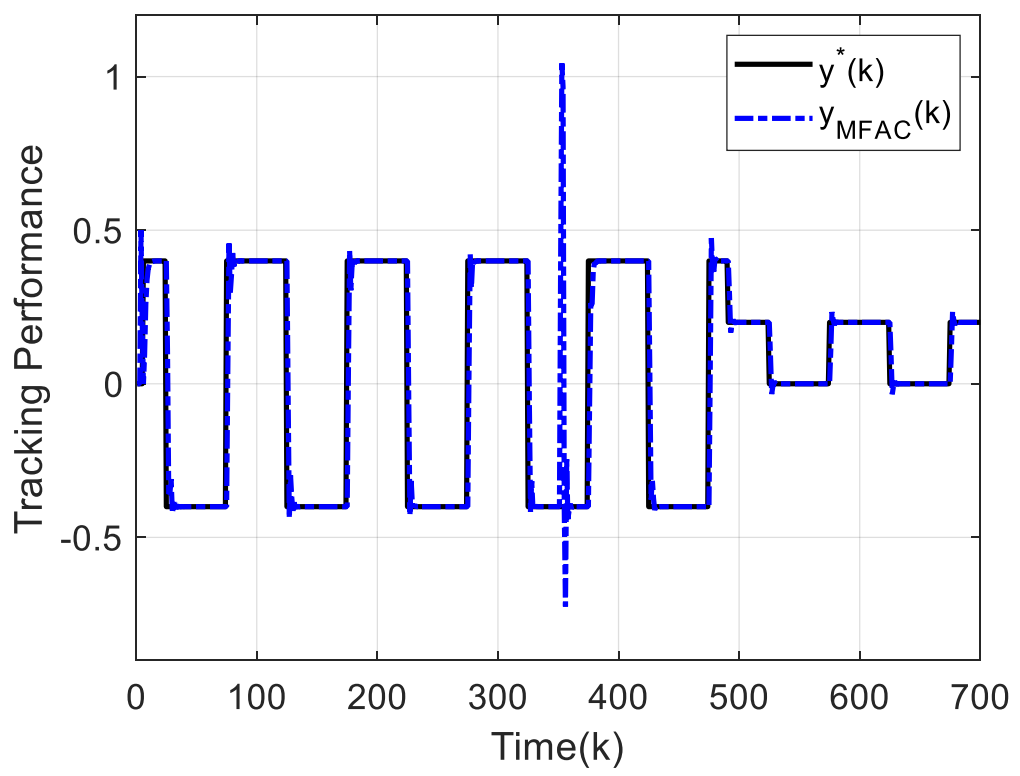

Fig. 4 Tracking performance

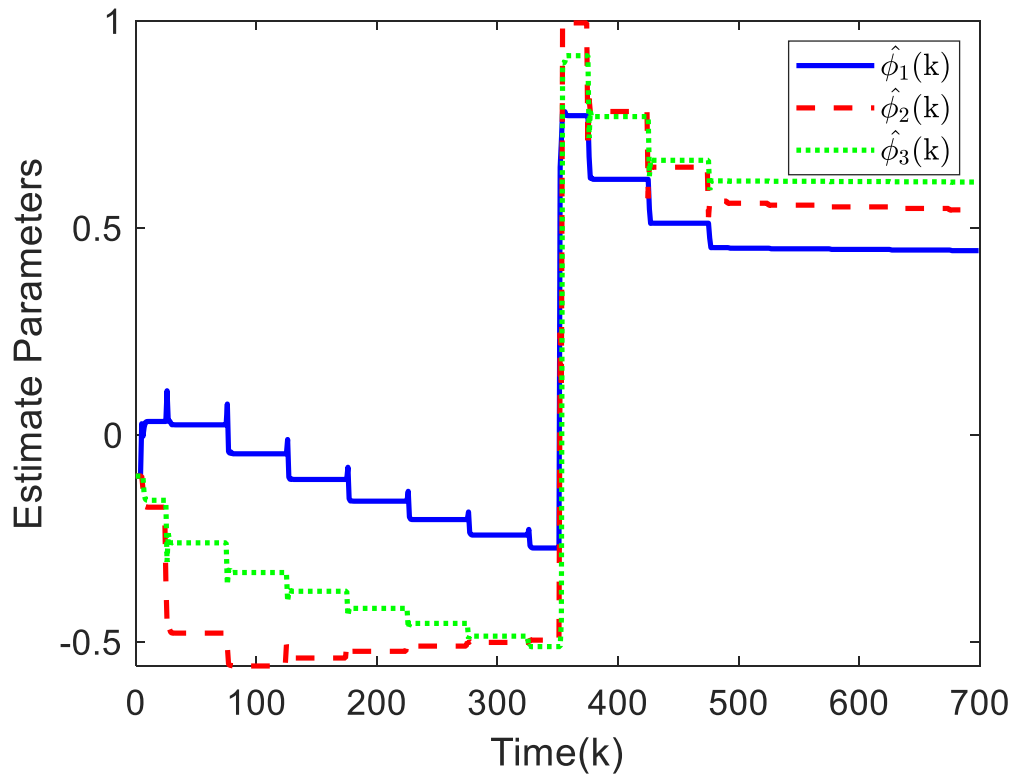

Fig. 5 Elements in estimated PG vector

From Fig. 3 and Fig. 5, we can see the sign of the estimated leading coefficient of control input $\hat{\phi}_{Ly+1}(k)$ changes at time 350. If the $\hat{\phi}_{Ly+1}(k)$ is reset according to [1]-[3], the sign of $\hat{\phi}_{Ly+1}(k)$ will become opposite to the actual value after the time of 350. In this way, EDLM can never objectively describe the actual system model, leading to the failure of the controller Therefore, we allow the estimation algorithm to continue operating in its own way without resetting its value, aiming to validate the fact that all signs of the estimated elements of PG are capable of changing and that the sign of the estimated leading coefficient $\hat{\phi}_{Ly+1}(k)$ of the control input should be altered according to the actual system model. However, the sign of $\hat{\phi}_{Ly+1}(k)$ unchanged is an essential precondition for the current stability analysis through the contraction mapping technique. Therefore, it will be more reasonable for us to analyze the stability of the system through the closed-loop function (11) and the static error.

*Example* 2: In this example, the following discrete-time SISO nonlinear system is considered.

$$y(k+1) = -y^2(k) + u(k) \qquad 1 \le k \le 201 \tag{15}$$

We choose the desired trajectory with

$$y^*(k) = -k^2, \qquad 1 \le k \le 201 \tag{16}$$

to validate the conclusion above about the steady-state error. The MFAC controller (8) is designed with $\hat{\phi}_2(k) = 1$ and $\hat{\phi}_1(k) = \sum_{i=1}^{2} \frac{1}{i!} \frac{\partial^i f(\varphi(k-1))}{\partial y^i(k-1)} \Delta y^{i-1}(k) = -2y(k-1) - \Delta y(k)$ . The outputs of the system controlled by MFAC with $\lambda$=0, $\lambda$=1×10$^{-5}$ and $\lambda$=3×10$^{-5}$ are shown in Fig. 6. The calculated PG vector is shown in Fig. 7

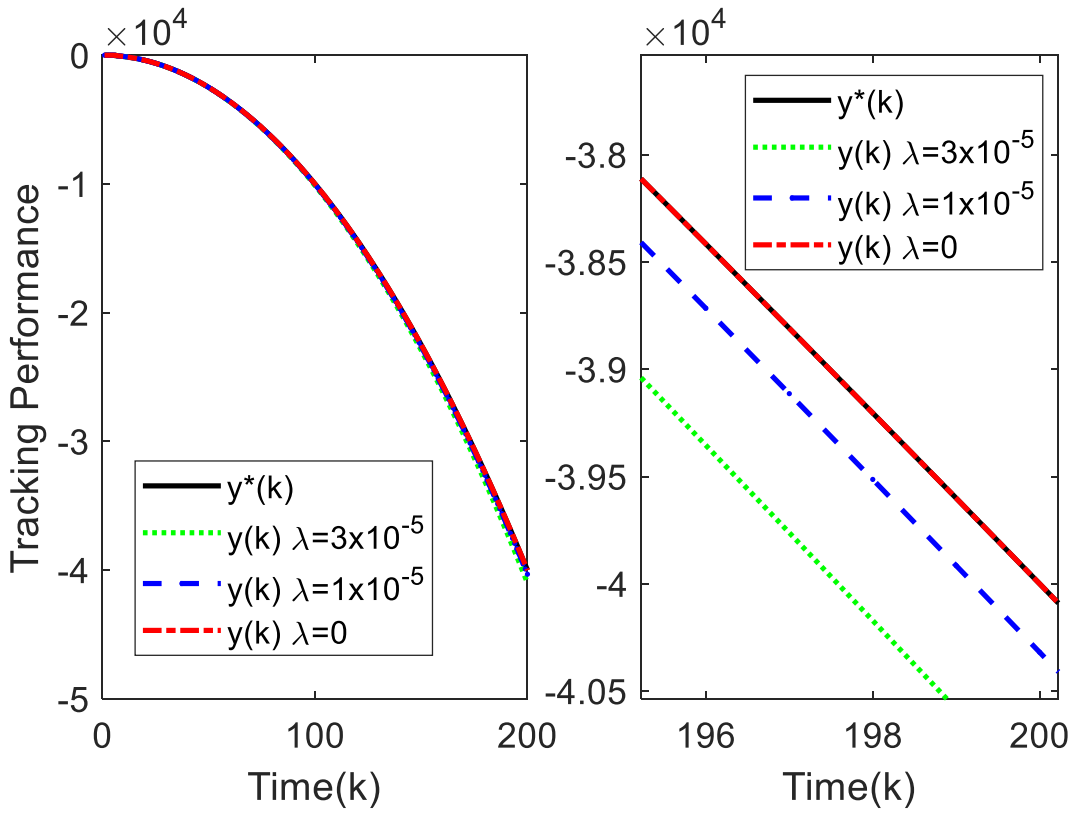

Fig. 6 Tracking performance

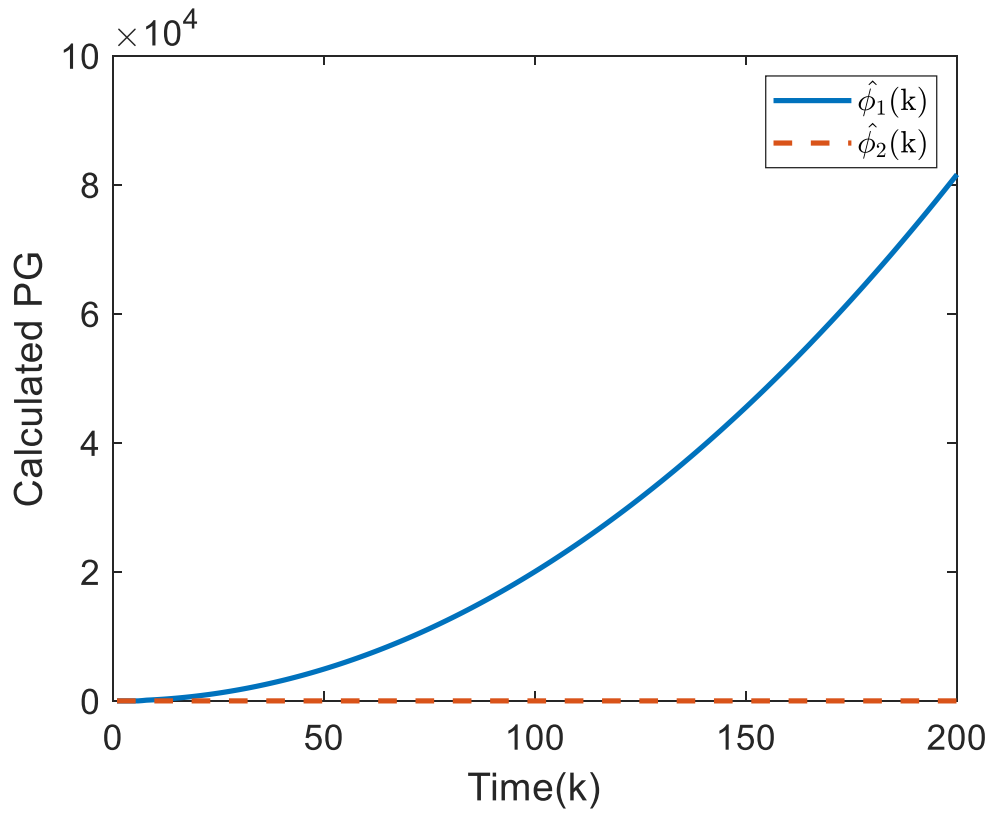

Fig. 7 Calculated PG vector

From Fig. 6, it is straightforward for someone to see that the static error will increase by raising $\lambda$. When $\lambda$=0, $y$(200)=$y^*$(200)=−4×10$^4$. Furthermore, we can conclude that the tracking error concerning the desired trajectory $k^n$ ( $n = 1, 2, \cdots$ ) will not converge to zero until $\lambda$=0. This fact contradicts [1]-[3] which showed that the tracking error of the system controlled by MFAC converges to zero under the condition that $\lambda$ is sufficiently large.

We change the desired trajectory (16) into

$$y^*(k) = -k, \qquad 1 \le k \le 500 \tag{17}$$

and choose $\lambda$=0.001. Define $e(k)$ as the tracking error at time $k$ and $E(k\text{-}Inf)$ as a property of transient tendency, i.e., "steady-state error" of the nonlinear system model at time $k$ which is calculated by (12), i.e.,

$$E(k - Inf) = \lim_{\substack{z \to 1 \\ \lambda = 10^{-3}}} \frac{z-1}{z} \frac{\lambda(1-z^{-1})[1-\phi_1(k)]}{\lambda(1-z^{-1})(1-\phi_1(k)) + \phi_2(k)} \frac{z}{(z-1)^2} \tag{18}$$

Fig. 8 shows the tracking performance and calculated PG. Fig. 9 shows the comparison between $E(k\text{-}Inf)$ and $e(k)$, $c(k)$= $E(k\text{-}Inf)/e(k)$ and $e(k)$-$E(k\text{-}Inf)$.

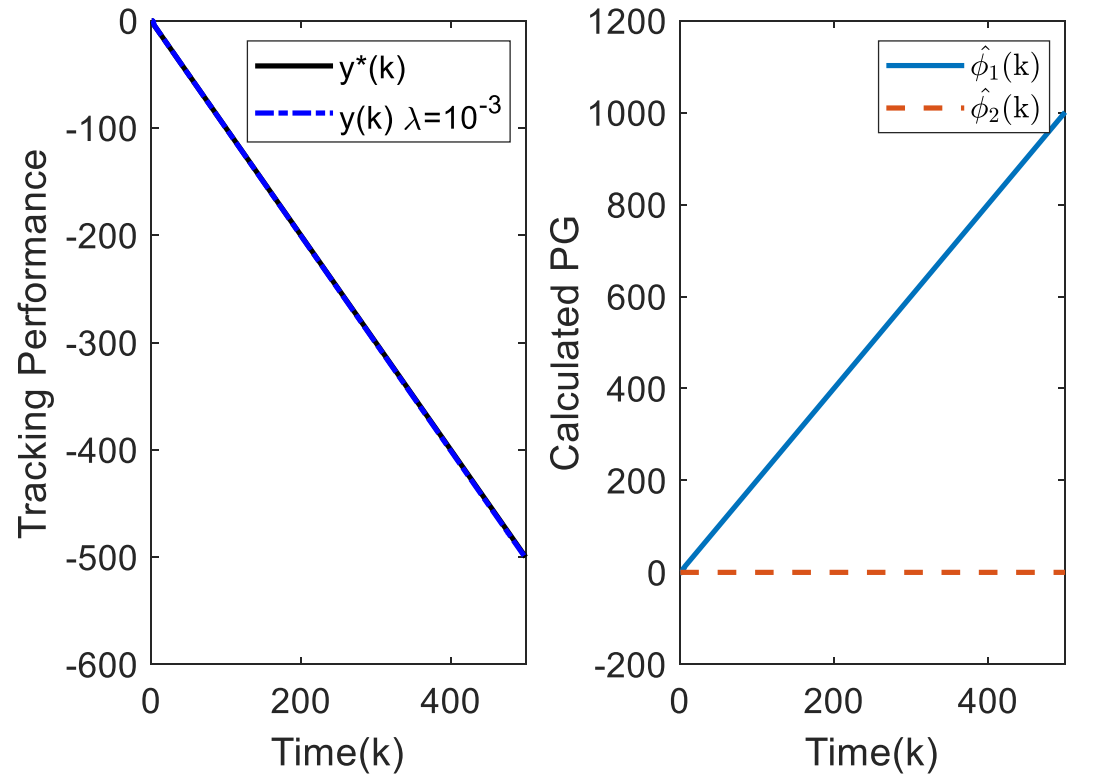

Fig. 8 Tracking performance and Calculated PG

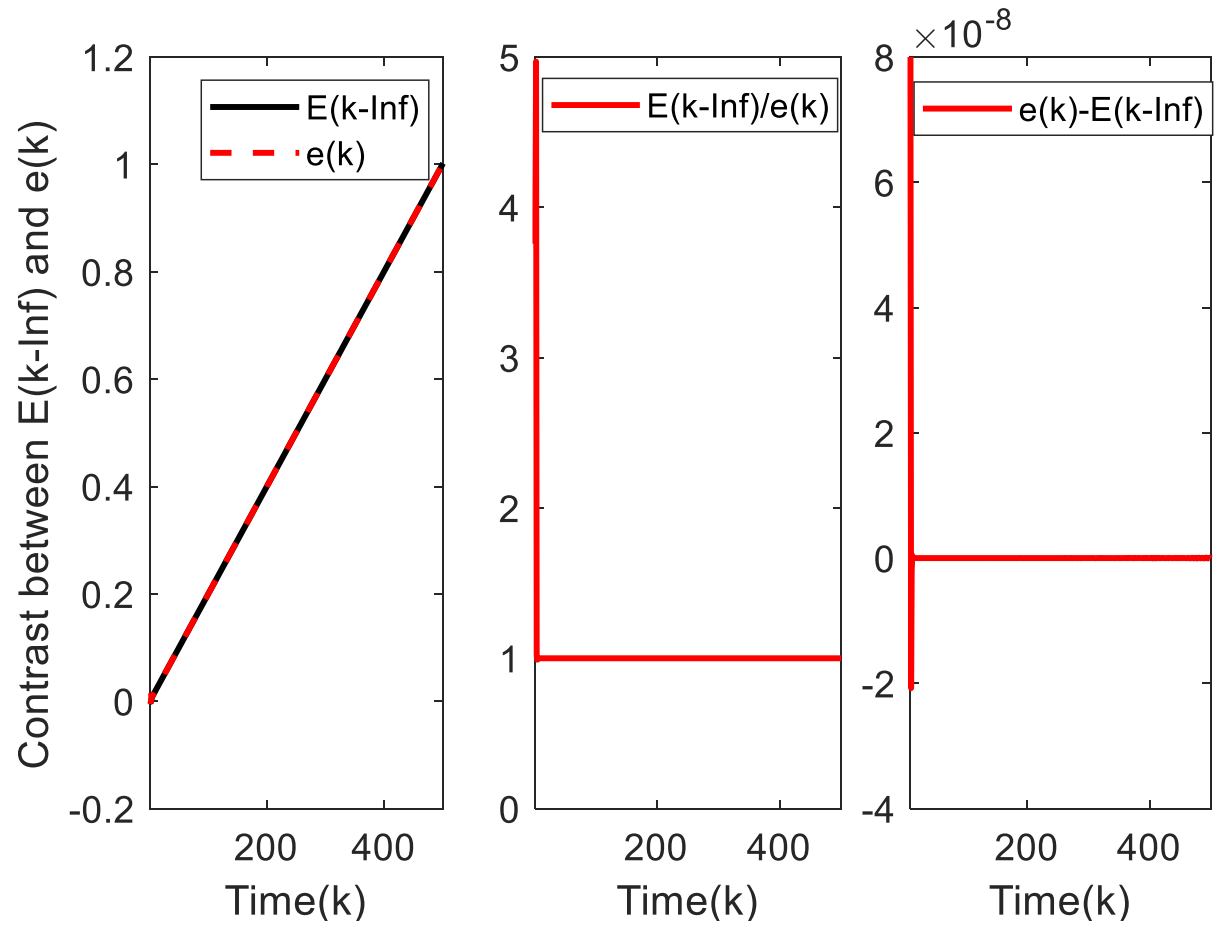

Fig. 9 Contrast between $E(k\text{-}Inf)$ and $e(k)$

Fig. 9 shows that $e(k)$ is very close to $E(k\text{-}Inf)$, confirming our conclusions about transient characteristics.

## III. Model-Free Adaptive Control For Nonlinear Systems

### A. *System* (1) *with slight nonlinearity*

To explore the application of MFAC in nonlinear systems further, we should begin with Case 2 in the Appendix. The system orders $n_y$+1 and $n_u$+1 are supposed to be known, and we let $L_y$=$n_y$+1 and $L_u$=$n_u$+1. Since $f(\cdots)$ in (1) is differentiable at the point $[y(k-1), \cdots, y(k-n_y-1),\ u(k-1), \cdots, u(k-n_u-1)]$ , the change in function $y(k) = f(y(k-1), \cdots, y(k-n_y-1), u(k-1), \cdots, u(k-n_u-1))$ as the variables change from $[y(k-1),\cdots,y(k-n_y-1),\ u(k-1),\cdots,u(k-n_u-1)]$ to $[y(k-1)+dy(k),\cdots,y(k-n_y-1)+dy(k-ny),u(k-1)+du(k),\cdots,u(k-n_u-1)+du(k-n_u)]$ is approximated by the total differential $dy(k+1)$ as shown in (19).

$$\Delta y(k+1)\approx dy(k+1)=\frac{\partial f(\boldsymbol{\varphi}(k-1))}{\partial y(k-1)}dy(k)+\cdots+\frac{\partial f(\boldsymbol{\varphi}(k-1))}{\partial y(k-n_y-1)}dy(k-n_y)+\frac{\partial f(\boldsymbol{\varphi}(k-1))}{\partial u(k-1)}du(k)+\cdots+\frac{\partial f(\boldsymbol{\varphi}(k-1))}{\partial u(k-n_u-1)}du(k-n_u) \tag{19}$$

The approximation $\Delta y(k+1)\approx dy(k+1)$ improves as $dy(k),\cdots,dy(k-n_y),\ du(k),\cdots,du(k-n_u)$ approach 0. Then we make a further approximation to have model (20), which is also discussed in the Appendix.

$$\begin{aligned}\Delta y(k+1)&=\boldsymbol{\phi}_L^T(k)\Delta\boldsymbol{H}(k)\\&=\frac{\partial f(\boldsymbol{\varphi}(k-1))}{\partial y(k-1)}\Delta y(k)+\cdots+\frac{\partial f(\boldsymbol{\varphi}(k-1))}{\partial y(k-n_y-1)}\Delta y(k-n_y)\\&\quad+\frac{\partial f(\boldsymbol{\varphi}(k-1))}{\partial u(k-1)}\Delta u(k)+\cdots+\frac{\partial f(\boldsymbol{\varphi}(k-1))}{\partial u(k-n_u-1)}\Delta u(k-n_u)\end{aligned} \tag{20}$$

where
$\Delta\boldsymbol{H}(k)=[\Delta y(k),\cdots,\Delta y(k-n_y),\ \Delta u(k),\cdots,\Delta u(k-n_u)]^T$,

$$\boldsymbol{\phi}_L(k)=\begin{bmatrix}\boldsymbol{\phi}_{Ly}(k)\\ \boldsymbol{\phi}_{Lu}(k)\end{bmatrix}=[\phi_1(k),\cdots,\phi_{Ly}(k),\phi_{Ly+1}(k),\cdots,\phi_{Ly+Lu}(k)]^T$$
$$=[\frac{\partial f(\boldsymbol{\varphi}(k-1))}{\partial y(k-1)},\cdots,\frac{\partial f(\boldsymbol{\varphi}(k-1))}{\partial y(k-n_y-1)},\frac{\partial f(\boldsymbol{\varphi}(k-1))}{\partial u(k-1)},\cdots,\frac{\partial f(\boldsymbol{\varphi}(k-1))}{\partial u(k-n_u-1)}]^T$$

When the control period of the system is sufficiently small, the model (20) accurately describes the system. Similarly to Section II, the MFAC controller is still written as (8).

On the other hand, we can design the MFAC controller by minimizing the quadratic function (9) subject to constraint $u_m(k)\le u(k)\le u_M(k)$, similar to the following *B* part.

*B. System* (1) *with strong nonlinearity*

i) If the system is strongly nonlinear and the control period is not sufficiently small, the obtained $\boldsymbol{\phi}_L(k)$ may have an apparent change from time *k* to *k*+1, sometimes leading to poor system performance. To this end, we recommend applying the iterative MFAC controller, as suggested in [8] and [9]. The controller will be

$$\begin{aligned}\Delta u(k+i\mid k)&=\frac{\dfrac{\partial f(\boldsymbol{\varphi}(k-1+i\mid k))}{\partial u(k-1+i\mid k)}}{\lambda(k+i,k)+\left[\dfrac{\partial f(\boldsymbol{\varphi}(k-1+i\mid k))}{\partial u(k-1+i\mid k)}\right]^2}[y^*(k+1)\\&\quad-y(k+i\mid k)-\sum_{j=1}^{ny+1}\frac{\partial f(\boldsymbol{\varphi}(k-1+i\mid k))}{\partial y(k-j+i\mid k)}\Delta y(k-j+1+i\mid k)\\&\quad-\sum_{j=2}^{nu+1}\frac{\partial f(\boldsymbol{\varphi}(k-1+i\mid k))}{\partial u(k-j+i\mid k)}\Delta u(k-j+1+i\mid k)]\end{aligned} \tag{21}$$

where *i* is the iteration number before the control input is sent to the system at *k*+1.

ii) On the other hand, if the function $f(\cdots)$ has derivatives of all orders at any operating points, the system model (1) can be described as (22) by the Appendix.

$$\begin{aligned}\Delta y(k+1)&=\frac{\partial f(\boldsymbol{\varphi}(k-1))}{\partial y(k-1)}\Delta y(k)+\cdots+\frac{\partial f(\boldsymbol{\varphi}(k-1))}{\partial y(k-n_y-1)}\Delta y(k-n_y)\\&\quad+\frac{\partial f(\boldsymbol{\varphi}(k-1))}{\partial u(k-1)}\Delta u(k)+\cdots+\frac{\partial f(\boldsymbol{\varphi}(k-1))}{\partial u(k-n_u-1)}\Delta u(k-n_u)\\&\quad+\gamma(k)\end{aligned} \tag{22}$$

where $\boldsymbol{\gamma}(k)$ is defined in the Appendix. Let

$$\begin{aligned}\boldsymbol{\phi}_L(k)=[&\frac{\partial f(\boldsymbol{\varphi}(k-1))}{\partial y(k-1)}+\varepsilon_1(k),\cdots,\frac{\partial f(\boldsymbol{\varphi}(k-1))}{\partial y(k-n_y-1)}+\varepsilon_{Ly}(k),\\&\frac{\partial f(\boldsymbol{\varphi}(k-1))}{\partial u(k-1)}+\varepsilon_{Ly+1}(k),\cdots,\frac{\partial f(\boldsymbol{\varphi}(k-1))}{\partial u(k-n_u-1)}+\varepsilon_{Ly+Lu}(k)]^T\end{aligned} \tag{23}$$

$$\begin{aligned}\varepsilon_{i+1}(k)&=\frac{1}{2!}\frac{\partial^2 f(\boldsymbol{\varphi}(k-1))}{\partial y^2(k-i-1)}\Delta y(k-i)+\frac{1}{3!}\frac{\partial^3 f(\boldsymbol{\varphi}(k-1))}{\partial y^3(k-i-1)}\Delta y^2(k-i)\\&\quad+\frac{1}{4!}\frac{\partial^4 f(\boldsymbol{\varphi}(k-1))}{\partial y^4(k-i-1)}\Delta y^3(k-i)+\cdots\end{aligned} \tag{24}$$

$$\begin{aligned}\varepsilon_{Ly+1+j}(k)&=\frac{1}{2!}\frac{\partial^2 f(\boldsymbol{\varphi}(k-1))}{\partial u^2(k-j-1)}\Delta u(k-j)+\frac{1}{3!}\frac{\partial^3 f(\boldsymbol{\varphi}(k-1))}{\partial u^3(k-j-1)}\\&\quad\bullet\Delta u^2(k-j)+\frac{1}{4!}\frac{\partial^4 f(\boldsymbol{\varphi}(k-1))}{\partial u^4(k-j-1)}\Delta u^3(k-j)+\cdots\end{aligned} \tag{25}$$

, $i=0,\cdots,n_y$ and $j=0,\cdots,n_u$, where (24) and (25) are collected from (41) in Appendix, and then (22) is rewritten as (4).

a) If $\frac{\partial f(\boldsymbol{\varphi}(k-1))}{\partial u(k-1)}$ is a nonzero constant, the controller is still described by (8) and the closed-loop system equation at time *k* is still (10).

b) If there exists $\frac{1}{n!}\frac{\partial^n f(\boldsymbol{\varphi}(k-1))}{\partial u^n(k-1)}\neq 0$ ($n\ge 1$) and $\frac{1}{j!}\frac{\partial^j f(\boldsymbol{\varphi}(k-1))}{\partial u^j(k-1)}=0$ ($j>n$), we may obtain the control law by minimizing the following cost function of at least 2*n*-th degree (26) or by minimizing the cost function (27) subject to constraint $u_m(k)\le u(k)\le u_M(k)$.

$$J=\left|y^*(k+1)-y(k+1)\right|^2+\lambda\left|\Delta u(k)\right|^2$$
$$=\left|y^*(k+1)-y(k)-\sum_{i=1}^{Ly}\phi_i(k)\Delta y(k-i+1)\right. \tag{26}$$
$$\left.-\sum_{i=Ly+1}^{Ly+Lu}\phi_i(k)\Delta u(k+L_y-i+1)\right|^2+\lambda\left|\Delta u(k)\right|^2$$

$$\min_{u(k)\ \text{s.t.}\ u_m(k)\le u(k)\le u_M(k)} J=\left|y^*(k+1)-y(k+1)\right|^2+\lambda\left|\Delta u(k)\right|^2 \tag{27}$$

Besides, to obtain the controller shown as (8) or to simplify the minimization of (27), we may simplify the controller design process through an approximation:

$$\phi_{Ly+1}(k)\approx\frac{\partial f(\boldsymbol{\varphi}(k-1))}{\partial u(k-1)}+\frac{1}{2!}\frac{\partial^2 f(\boldsymbol{\varphi}(k-1))}{\partial u^2(k-1)}\Delta u(k-1)$$
$$+\frac{1}{3!}\frac{\partial^3 f(\boldsymbol{\varphi}(k-1))}{\partial u^3(k-1)}\Delta u^2(k-1)+\cdots \tag{28}$$

where the coefficient (28) stems from (41) and $\Delta u(k)$ is replaced by $\Delta u(k-1)$.

*Simulations:*

*Example* 3: The following discrete-time SISO nonlinear system is considered.

$$y(k+1)=0.2y^2(k)+2u(k)+u^2(k)+2u^5(k-1)$$
$$+\cos(u(k-1))+u^6(k-2) \tag{29}$$

The desired output trajectory is

$$y^*(k)=\begin{cases}0.5\sin(k/50)+0.5\cos(k/3) & \\ \quad+0.5\sin(k/10) & 1\le k\le 350\\ 0.3+0.3\times(-1)^{round(k/50)} & 351\le k\le 700\end{cases} \tag{30}$$

The controller parameters and initial values for MFAC are listed in Table II.

TABLE II Parameter Settings for MFAC

| Parameter | MFAC (6) |
|---|---|
| Order | $L_y$=1, $L_u$=3 |
| $\lambda$ | 1.5 |
| Initial value | |
| $\hat{\phi}_L(3)=\hat{\phi}_L(2)=\hat{\phi}_L(1)$ | [0.01, 0.01, 0.01,0.01] |
| $y(0:4)$, $u(0:3)$ | 0 |

The controller MFAC 1 is designed by minimizing a quartic equation (26) shown as

$$J=\left|y^*(k+1)-y(k+1)\right|^2+\lambda\left|\Delta u(k)\right|^2$$
$$=\left|y^*(k+1)-y(k)-\hat{\phi}_1(k)\Delta y(k)-\hat{\phi}_3(k)\Delta u(k-1)-\hat{\phi}_4(k)\Delta u(k-2)\right.$$
$$\left.-\frac{\partial f(\boldsymbol{\varphi}(k-1))}{\partial u(k-1)}\Delta u(k)-\frac{1}{2!}\frac{\partial^2 f(\boldsymbol{\varphi}(k-1))}{\partial u^2(k-1)}\Delta u^2(k)\right|^2+\lambda\left|\Delta u(k)\right|^2 \tag{31}$$

with $\hat{\phi}_1(k)=\sum_{i=1}^{2}\frac{1}{i!}\frac{\partial^i f(\boldsymbol{\varphi}(k-1))}{\partial y^i(k-1)}\Delta y^{i-1}(k)$,

$$\hat{\phi}_2(k)=\sum_{i=1}^{2}\frac{1}{i!}\frac{\partial^i f(\boldsymbol{\varphi}(k-1))}{\partial u^i(k-1)}\Delta u^{i-1}(k)=2+2u(k-1)+\Delta u(k),$$

$$\hat{\phi}_3(k)=\sum_{i=1}^{5}\frac{1}{i!}\frac{\partial^i f(\boldsymbol{\varphi}(k-1))}{\partial u^i(k-2)}\Delta u^{i-1}(k-1),$$

$$\hat{\phi}_4(k)=\sum_{i=1}^{6}\frac{1}{i!}\frac{\partial^i f(\boldsymbol{\varphi}(k-1))}{\partial u^i(k-3)}\Delta u^{i-1}(k-2).$$

We apply the controller (8) named MFAC 2. $\hat{\phi}_1(k)$, $\hat{\phi}_3(k)$ and $\hat{\phi}_4(k)$ are the same as those of MFAC 1 and $\hat{\phi}_2(k)$ is approximated by (28): $\hat{\phi}_2(k)=2+2u(k-1)+\Delta u(k-1)$.

The controller MFAC 3 is designed to minimize (24) subject to the control input constraint $-0.6\le u(k)\le -0.2$

Fig. 10 shows the tracking performance of the system controlled by MFAC 1, MFAC 2 and MFAC 3. Fig. 11 shows the control input. Fig. 12 shows the elements in the calculated PG vector of MFAC 1.

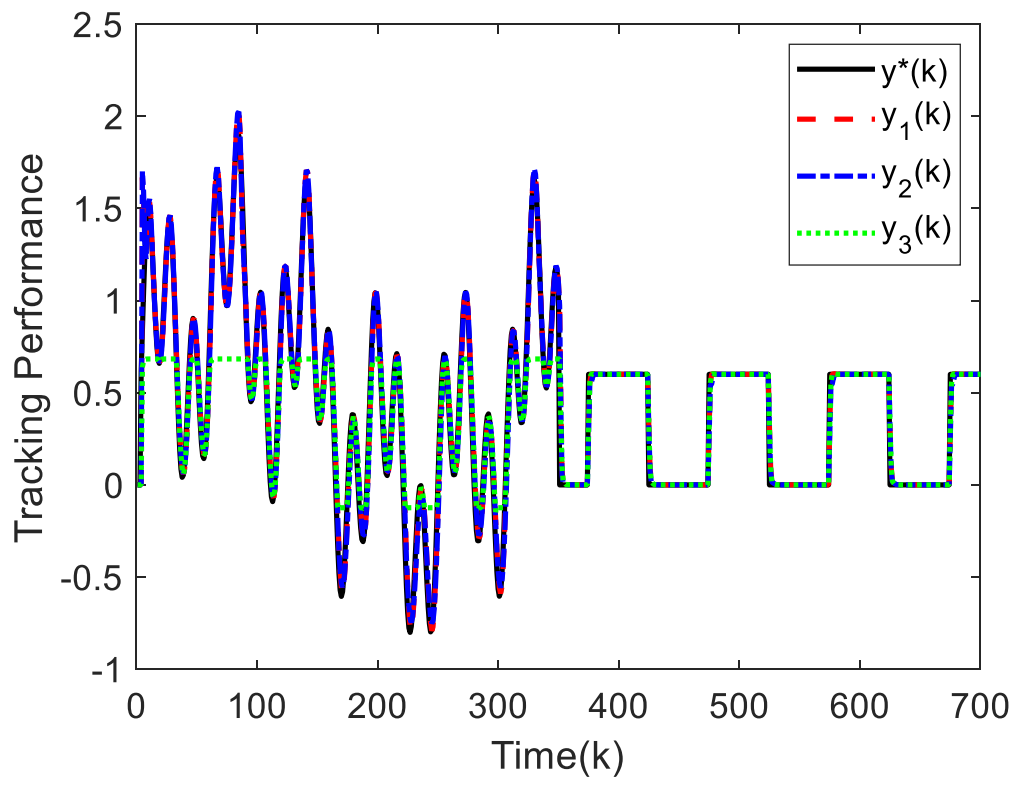

Fig. 10 Tracking performance

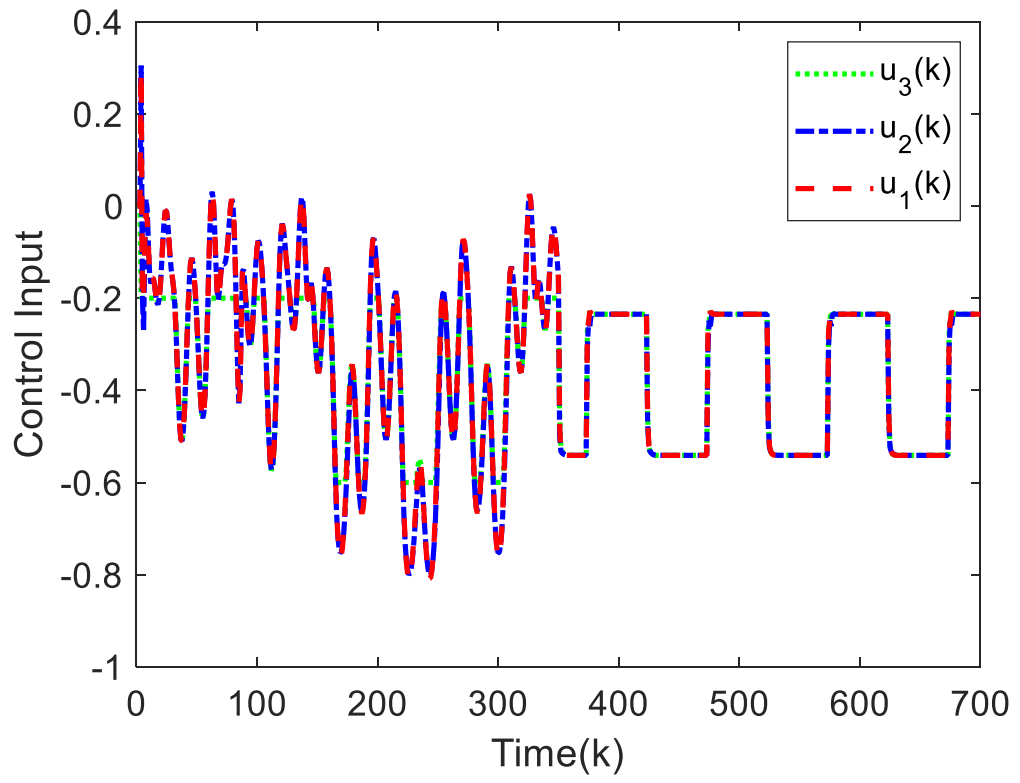

Fig. 11 Control input

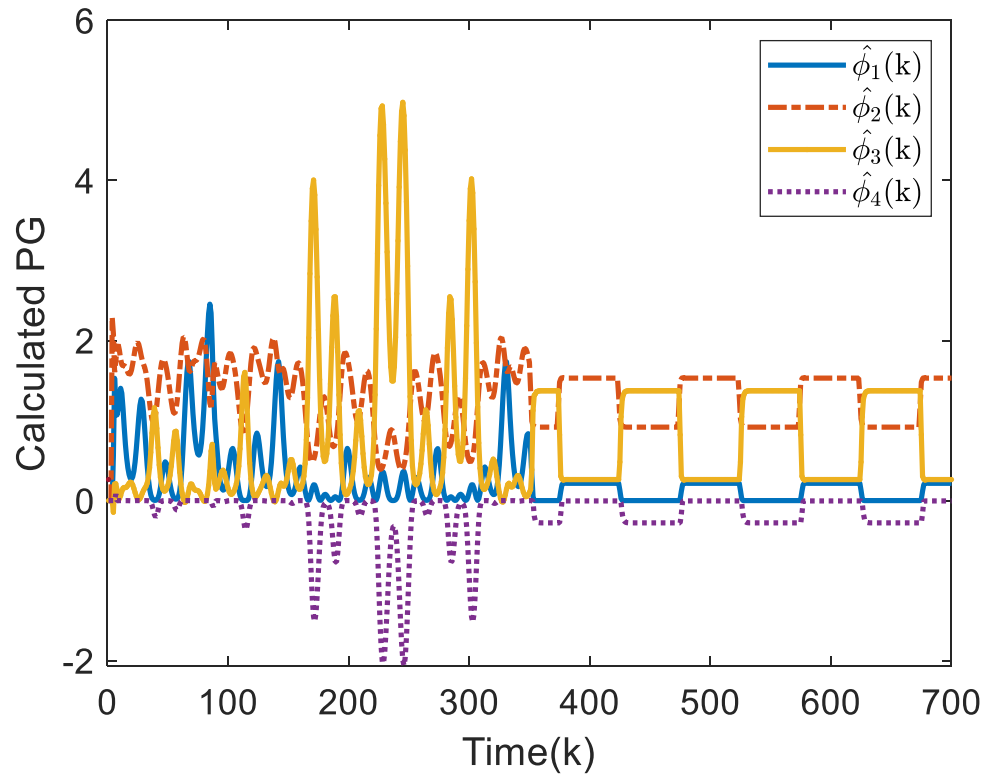

Fig. 12 Elements in calculated PG vector

Question: Why don't we use the MFAC by introducing an online estimation algorithm like [1]?

Answer: Because the model can hardly be established in this way. One can refer to the Remark in [7] for detailed explanations.

## IV. CONCLUSION

In this note, the stability of the system and the chosen parameter $\lambda$ are analyzed by the closed-loop function, and we have figured out that the current MFAC methods are not analyzed correctly. Then, several simulated examples are used to validate the viewpoints. According to its nature, we should rename MFAC as the incremental one-step-ahead optimal controller.

## APPENDIX

Proof of *Theorem 1*

*Proof*: Case 1: $1\le L_y \le n_y$ and $1\le L_u \le n_u$

From (1), we have

$$\begin{aligned}&\Delta y(k+1)=\\&f(y(k),\cdots,y(k-L_y+1),y(k-L_y)\cdots,y(k-n_y),u(k),\\&\quad\cdots,u(k-L_u+1),u(k-L_u),\cdots,u(k-n_u))\\&-f(y(k-1),\cdots,y(k-L_y),y(k-L_y),\cdots,y(k-n_y),u(k-1),\\&\quad\cdots,u(k-L_u),u(k-L_u),\cdots,u(k-n_u))\\&+f(y(k-1),\cdots,y(k-L_y),y(k-L_y),\cdots,y(k-n_y),u(k-1),\\&\quad\cdots,u(k-L_u),u(k-L_u),\cdots,u(k-n_u))\\&-f(y(k-1),\cdots,y(k-L_y),y(k-L_y-1),\cdots,y(k-n_y-1),\\&\quad u(k-1),\cdots,u(k-L_u),u(k-L_u-1),\cdots,u(k-n_u-1))\end{aligned} \tag{32}$$

On the basis of *Assumption 1* and the definition of differentiability in [15], (32) becomes

$$\begin{aligned}\Delta y(k+1)&=\frac{\partial f(\boldsymbol{\varphi}(k-1))}{\partial y(k-1)}\Delta y(k)+\cdots+\frac{\partial f(\boldsymbol{\varphi}(k-1))}{\partial y(k-L_y)}\Delta y(k-L_y+1)\\&+\frac{\partial f(\boldsymbol{\varphi}(k-1))}{\partial u(k-1)}\Delta u(k)+\cdots+\frac{\partial f(\boldsymbol{\varphi}(k-1))}{\partial u(k-L_u)}\Delta u(k-L_u+1)\\&+\varepsilon_1(k)\Delta y(k)+\cdots+\varepsilon_{Ly}(k)\Delta y(k-L_y+1)+\varepsilon_{Ly+1}(k)\Delta u(k)\\&+\cdots+\varepsilon_{Ly+Lu}(k)\Delta u(k-L_u+1)+\psi(k)\end{aligned} \tag{33}$$

where

$$\begin{aligned}\psi(k)\triangleq\ &f(y(k-1),\cdots,y(k-L_y),y(k-L_y),\cdots,y(k-n_y),\\&u(k-1),\cdots,u(k-L_u),u(k-L_u),\cdots,u(k-n_u))\\&-f(y(k-1),\cdots,y(k-L_y),y(k-L_y-1),\cdots,y(k-n_y-1),\\&u(k-1),\cdots,u(k-L_u),u(k-L_u-1),\cdots,u(k-n_u-1))\end{aligned} \tag{34}$$

$\frac{\partial f(\boldsymbol{\varphi}(k-1))}{\partial y(k-i-1)}$, $0\le i\le L_y-1$ and $\frac{\partial f(\boldsymbol{\varphi}(k-1))}{\partial u(k-j-1)}$, $0\le j\le L_u-1$ denote the partial derivative values of $f(\boldsymbol{\varphi}(k-1))$ with respect to the ($i$+1)-th variable and the ($n_y+2+j$)-th variable, respectively. And $\varepsilon_1(k),\cdots,\varepsilon_{Ly+Lu}(k)$ are functions that depend only on $\Delta y(k),\cdots,\Delta y(k-L_y+1),\Delta u(k),\cdots,\Delta u(k-L_u+1)$, with $(\varepsilon_1(k),\cdots,\varepsilon_{Ly+Lu}(k))\to(0,\cdots,0)$ when $(\Delta y(k),\cdots,\Delta y(k-L_y+1),\Delta u(k),\cdots,\Delta u(k-L_u+1))\to(0,\cdots,0)$. This also implies that $(\varepsilon_1(k),\cdots,\varepsilon_{Ly+Lu}(k))$ can be regarded as $(0,\cdots,0)$ when the control period of the system is sufficiently small.

We consider the following equation with the vector $\boldsymbol{\eta}(k)$ for each time $k$:

$$\psi(k)=\boldsymbol{\eta}^T(k)\Delta\boldsymbol{H}(k) \tag{35}$$

Owing to $\|\Delta\boldsymbol{H}(k)\|\ne 0$, (35) must have at least one solution $\boldsymbol{\eta}^*(k)$. Let

$$\begin{aligned}\boldsymbol{\phi}_L(k)=\boldsymbol{\eta}^*(k)+[&\frac{\partial f(\boldsymbol{\varphi}(k-1))}{\partial y(k-1)}+\varepsilon_1(k),\cdots,\frac{\partial f(\boldsymbol{\varphi}(k-1))}{\partial y(k-L_y)}+\varepsilon_{Ly}(k),\\&\frac{\partial f(\boldsymbol{\varphi}(k-1))}{\partial u(k-1)}+\varepsilon_{Ly+1}(k),\cdots,\frac{\partial f(\boldsymbol{\varphi}(k-1))}{\partial u(k-L_u)}+\varepsilon_{Ly+Lu}(k)]^T\end{aligned} \tag{36}$$

(33) can be described as follows:

$$\Delta y(k+1)=\boldsymbol{\phi}_L^T(k)\Delta\boldsymbol{H}(k) \tag{37}$$

Case 2: $L_y=n_y+1$ and $L_u=n_u+1$

On the basis of *Assumption 1* and the definition of differentiability in [15], (1) becomes

$$\begin{aligned}\Delta y(k+1)&=\frac{\partial f(\boldsymbol{\varphi}(k-1))}{\partial y(k-1)}\Delta y(k)+\cdots+\frac{\partial f(\boldsymbol{\varphi}(k-1))}{\partial y(k-n_y-1)}\Delta y(k-n_y)\\&+\frac{\partial f(\boldsymbol{\varphi}(k-1))}{\partial u(k-1)}\Delta u(k)+\cdots+\frac{\partial f(\boldsymbol{\varphi}(k))}{\partial u(k-n_u-1)}\Delta u(k-n_u)\\&+\gamma(k)\end{aligned} \tag{38}$$

where

$$\begin{aligned}\gamma(k)=\ &\varepsilon_1(k)\Delta y(k)+\cdots+\varepsilon_{Ly}(k)\Delta y(k-n_y)\\&+\varepsilon_{Ly+1}(k)\Delta u(k)+\cdots+\varepsilon_{Ly+Lu}(k)\Delta u(k-n_u)\end{aligned} \tag{39}$$

We let

$$\begin{aligned}\boldsymbol{\phi}_L(k)=[&\frac{\partial f(\boldsymbol{\varphi}(k-1))}{\partial y(k-1)}+\varepsilon_1(k),\cdots,\frac{\partial f(\boldsymbol{\varphi}(k-1))}{\partial y(k-n_y-1)}+\varepsilon_{Ly}(k),\\&\frac{\partial f(\boldsymbol{\varphi}(k-1))}{\partial u(k-1)}+\varepsilon_{Ly+1}(k),\cdots,\frac{\partial f(\boldsymbol{\varphi}(k-1))}{\partial u(k-n_u-1)}+\varepsilon_{Ly+Lu}(k)]^T\end{aligned} \tag{40}$$

to rewrite (38) as (37), with $(\varepsilon_1(k),\cdots,\varepsilon_{Ly+Lu}(k))\to(0,\cdots,0)$ in nonlinear systems, when $(\Delta y(k),\cdots,\Delta y(k-n_y),\Delta u(k),\cdots,\Delta u(k-n_u))\to(0,\cdots,0)$. As to linear systems, we will always have $\boldsymbol{\phi}_L(k)=[\frac{\partial f(\boldsymbol{\varphi}(k-1))}{\partial y(k-1)},\cdots,\frac{\partial f(\boldsymbol{\varphi}(k-1))}{\partial y(k-n_y-1)},\frac{\partial f(\boldsymbol{\varphi}(k-1))}{\partial u(k-1)},\cdots,,\frac{\partial f(\boldsymbol{\varphi}(k-1))}{\partial u(k-n_u-1)}]^T$ no matter what $(\Delta y(k),\cdots,\Delta y(k-n_y),\Delta u(k),\cdots,\Delta u(k-n_u))$ is.

Additionally, if the function $f(\cdots)$ has derivatives of all orders on any operating points, we can obtain (41) by the Taylor series

$$\begin{aligned}\Delta y(k+1) = {} & [\Delta y(k)\frac{\partial}{\partial y(k-1)} + \cdots + \Delta y(k-n_y)\frac{\partial}{\partial y(k-n_y-1)} \\ & + \Delta u(k)\frac{\partial}{\partial u(k-1)} + \cdots + \Delta u(k-n_u)\frac{\partial}{\partial u(k-n_u-1)}] f(\boldsymbol{\varphi}(k-1)) \\ & + \frac{1}{2!}[\Delta y(k)\frac{\partial}{\partial y(k-1)} + \cdots + \Delta y(k-n_y)\frac{\partial}{\partial y(k-n_y-1)} \\ & + \Delta u(k)\frac{\partial}{\partial u(k-1)} + \cdots + \Delta u(k-n_u)\frac{\partial}{\partial u(k-n_u-1)}]^2 f(\boldsymbol{\varphi}(k-1)) \\ & + \cdots\end{aligned} \tag{41}$$

and then obtain a group of solutions (42), (43) for (39) as follows

$$\begin{aligned}\varepsilon_{i+1}(k) = {} & \frac{1}{2!}\frac{\partial^2 f(\boldsymbol{\varphi}(k-1))}{\partial y^2(k-i-1)}\Delta y(k-i) + \frac{1}{3!}\frac{\partial^3 f(\boldsymbol{\varphi}(k-1))}{\partial y^3(k-i-1)}\Delta y^2(k-i) \\ & + \frac{1}{4!}\frac{\partial^4 f(\boldsymbol{\varphi}(k-1))}{\partial y^4(k-i-1)}\Delta y^3(k-i) + \cdots\end{aligned} \tag{42}$$

$$\begin{aligned}\varepsilon_{Ly+1+j}(k) = {} & \frac{1}{2!}\frac{\partial^2 f(\boldsymbol{\varphi}(k-1))}{\partial u^2(k-j-1)}\Delta u(k-j) + \frac{1}{3!}\frac{\partial^3 f(\boldsymbol{\varphi}(k-1))}{\partial u^3(k-j-1)} \\ & \bullet\Delta u^2(k-j) + \frac{1}{4!}\frac{\partial^4 f(\boldsymbol{\varphi}(k-1))}{\partial u^4(k-j-1)}\Delta u^3(k-j) + \cdots\end{aligned} \tag{43}$$

, $i=0,\cdots,n_y$ and $j=0,\cdots,n_u$.

Case 3: $L_y>n_y+1$ and $L_u>n_u+1$

Based on Assumption 1 and the definition of differentiability in [15], (1) becomes

$$\begin{aligned}\Delta y(k+1) = {} & \frac{\partial f(\boldsymbol{\varphi}(k-1))}{\partial y(k-1)}\Delta y(k) + \cdots + \frac{\partial f(\boldsymbol{\varphi}(k-1))}{\partial y(k-n_y-1)}\Delta y(k-n_y) \\ & + \frac{\partial f(\boldsymbol{\varphi}(k-1))}{\partial u(k-1)}\Delta u(k) + \cdots + \frac{\partial f(\boldsymbol{\varphi}(k-1))}{\partial u(k-n_u-1)}\Delta u(k-n_u) \\ & + \varepsilon_1(k)\Delta y(k) + \cdots + \varepsilon_{n_y+1}(k)\Delta y(k-n_y) + \varepsilon_{Ly+1}(k)\Delta u(k) \\ & + \cdots + \varepsilon_{Ly+n_u+1}(k)\Delta u(k-n_u)\end{aligned} \tag{44}$$

Define

$$\begin{aligned}\gamma(k) = {} & \varepsilon_1\Delta y(k) + \cdots + \varepsilon_{n_y+1}\Delta y(k-n_y) + \varepsilon_{Ly+1}\Delta u(k) \\ & + \cdots + \varepsilon_{Ly+n_u+1}\Delta u(k-n_u)\end{aligned} \tag{45}$$

We consider the following equation with the vector $\boldsymbol{\eta}(k)$ for each time $k$:

$$\gamma(k) = \boldsymbol{\eta}^T(k)\Delta\boldsymbol{H}(k) \tag{46}$$

Owing to $\|\Delta\boldsymbol{H}(k)\| \neq 0$, (46) must have at least one solution $\boldsymbol{\eta}^*(k)$. Let

$$\begin{aligned}\boldsymbol{\phi}_L(k) = \boldsymbol{\eta}^*(k) + [ & \frac{\partial f(\boldsymbol{\varphi}(k-1))}{\partial y(k-1)}, \cdots, \frac{\partial f(\boldsymbol{\varphi}(k-1))}{\partial y(k-n_y-1)}, 0, \cdots, 0, \\ & \frac{\partial f(\boldsymbol{\varphi}(k-1))}{\partial u(k-1)}, \cdots, \frac{\partial f(\boldsymbol{\varphi}(k-1))}{\partial u(k-n_u-1)}, 0, \cdots, 0]^T\end{aligned} \tag{47}$$

Then (44) can be rewritten as (37).

Case 4: $L_y\geq n_y+1$ and $1\leq L_u<n_u+1$; $0\leq L_y<n_y+1$ and $L_u\geq n_u+1$.

The proof of Case 4 is similar to the above analysis process, we omit it.

We finished the proof of *Theorem 1*.

*Remark* 2: The UD in Case 2 ($L_y=n_y+1$ and $L_u=n_u+1$) is shown as follows.

(38) can be rewritten as

$$\begin{aligned}y(k) = {} & \frac{\partial f(\boldsymbol{\varphi}(k-1))}{\partial y(k-1)}y(k-1) + \cdots + \frac{\partial f(\boldsymbol{\varphi}(k-1))}{\partial y(k-n_y-1)}y(k-n_y-1) \\ & + \frac{\partial f(\boldsymbol{\varphi}(k-1))}{\partial u(k-1)}u(k-1) + \cdots + \frac{\partial f(\boldsymbol{\varphi}(k-1))}{\partial u(k-n_u-1)}u(k-n_u-1) + v(k)\end{aligned} \tag{48}$$

where (49) represents the UD around the operating point, according to the conception in [16].

$$\begin{aligned}v(k) = {} & y(k+1) - \frac{\partial f(\boldsymbol{\varphi}(k-1))}{\partial y(k-1)}y(k) - \cdots - \frac{\partial f(\boldsymbol{\varphi}(k-1))}{\partial y(k-n_y-1)}y(k-n_y) \\ & - \frac{\partial f(\boldsymbol{\varphi}(k-1))}{\partial u(k-1)}u(k) - \cdots - \frac{\partial f(\boldsymbol{\varphi}(k))}{\partial u(k-n_u-1)}u(k-n_u) - \gamma(k)\end{aligned} \tag{49}$$